\documentclass[twocolumn,showpacs,amsmath,amssymb,pra]{revtex4}
\usepackage{bbm}

\usepackage{graphicx}%
\usepackage{dcolumn}
\usepackage{amsmath}
\usepackage{bm}
\usepackage{color}
\newcommand{\be}{\begin{equation}}
\newcommand{\ee}{\end{equation}}
\newcommand{\bey}{\begin{eqnarray}}
\newcommand{\eey}{\end{eqnarray}}
\newcommand{\bw}{\begin{widetext}}
\newcommand{\ew}{\end{widetext}}

\newcommand{\ra}{\rangle}
\newcommand{\la}{\langle}
\newcommand{\br}{ {\bf r} }

\newcommand{\N}{{\cal{N} }}

\newcommand{\ba}{\begin{array}}
\newcommand{\ea}{\end{array}}
\newcommand{\bi}{\begin{itemize}}
\newcommand{\ei}{\end{itemize}}
\newcommand{\bem}{\begin{enumerate}}
\newcommand{\eem}{\end{enumerate}}

\begin{document}

 \title {Semiclassical approach to the quantum Loschmidt echo in deep quantum regions:
 from validity to breakdown
 }

 \author{Pinquan Qin, Qian Wang, and Wen-ge Wang\footnote{ Email address: wgwang@ustc.edu.cn}}

 \affiliation{
 Department of Modern Physics, University of Science and Technology of China,
 Hefei 230026, China
 }

 \date{\today}

 \begin{abstract}

 Semiclassical results are usually expected to be valid in the semiclassical regime.
 An interesting question is, in models in which appropriate effective Planck constants can be introduced,
 to what extent will a semiclassical prediction stay valid when the effective Planck constant is increased?
 In this paper, we numerically study this problem, focusing on  semiclassical predictions for the decay
 of the quantum Loschmidt echo in deep quantum regions.
 Our numerical simulations, carried out in the chaotic regime in the sawtooth model and
 in the kicked rotator model and also in the critical region of a 1D Ising chain in transverse field,
 show that the semiclassical predictions may work even in deep quantum regions,
 in particularly, for perturbation strength in the so-called Fermi-Golden-rule regime.

 \end{abstract}

 \maketitle


 \section{Introduction}

 The semiclassical theory is powerful in dealing with many problems in various fields of
 physics \cite{haa2001,ma1997}.
 It is usually expected to work in the semiclassical regime, in which the
 (effective) Planck constant is sufficiently small in a certain relative sense.
 An interesting question is, how deep in the quantum regime could the semiclassical predictions remain valid?
 Obviously, the answer should depend on the physical quantity of interest.

 In this paper, we study a quantity for which the semiclassical approach has recently
 been found to be quite successful in the semiclassical regime.
 It is the so-called quantum Loschmidt echo (LE) \cite{qc,pqc,per1984}, which is given by
 the overlap of the time evolution of the same initial state under two slightly different Hamiltonians,
 \begin{eqnarray} \nonumber
 M(t) & = & |m(t)|^2  \ \ \
 \\ \text{with} \ m(t) & = & \langle\Psi_0|\exp(iH_1t/\hbar) \exp(-iH_0t/\hbar) |\Psi_0\rangle,
 \label{le}
 \end{eqnarray}
 where $H_1 = H_0+\epsilon V$, with $\epsilon$ a small quantity and $V$ a generic perturbation.
 The quantity $m(t)$ is usually called the amplitude of the LE.
 The LE gives a measure to the stability of quantum motion under small perturbation.

 The LE has quite rich behaviors, depending on the nature of the dynamics of the underlying
 classical system, as well as on the perturbation strength.
 Usually, the LE has a quadratic decay within a certain initial time interval,
 as predicted by the first-order perturbation theory \cite{wis2003}.
 Beyond the initial time interval,
 in a chaotic system, loosely speaking, the LE has a Gaussian decay
 \cite{per1984,jac2001,pro2003,cer2002,pro2002t}
 below a perturbative border and has an exponential decay in the so-called
 Fermi-golden-rule (FGR) regime above the perturbative border with intermediate perturbation strength
 \cite{cer2002,pro2002t,pro2002,gut2010,wis2002,wisn2002,wgw2004}.
 With further increase of the perturbation strength,  in the so-called Lyapunov regime
 with relatively strong perturbation, the LE usually has a perturbation-independent  decay
 \cite{jal2001,sil2003,cuc2004,wgw2004,wgw2005,wgw2005t,gut2010};
 but, in certain cases, a perturbation-dependent oscillation in the decay rate may also appear
 \cite{wgw2004,wgw2005,wgw2005t,are2009,Gar2011}.

 On the other hand,
 in regular systems, in the case of one degree of freedom, the LE has a Gaussian decay \cite{pro2002},
 followed by a power-law decay \cite{JAB03,wan2007}.
 Meanwhile, in the case of many-degrees of freedom, the LE may have an exponential decay for
 times much shorter than the recurrence time of the LE \cite{wgw2010}.

 The above-discussed semiclassical predictions for the LE decay have been tested numerically
 in the deep semiclassical regime in some models,
 in which effective Planck constants can be suitably introduced.
 In this regime, the effective Planck constants are sufficiently small,
 such that the stationary phase approximation is applicable in the derivation of the semiclassical propagator
 from Feynman's path integral theory.
 Here, we are interested in the extent to which the predictions may remain valid in the opposite
 deep-quantum regime.
 In this regime, the effective Planck constants are not very small, such that the
 validity of the above-mentioned stationary phase approximation becomes questionable.
 To study this problem, it is necessary to rely mainly on numerical simulations in concrete models.
 Our numerical results obtained in the sawtooth model and in the kicked rotor model
 show that the semiclassical predictions may work well even in the deep quantum regime.

 We also study the LE decay in the vicinity of the quantum phase transition (QPT)
 in a one-dimensional(1D) Ising chain in transverse field.
 At a QPT, at which the ground level has level crossing with other level(s),
 the ground state has drastic change(s) in its fundamental properties \cite{sac1999}.
 Quantities borrowed from the quantum information field have been found useful in characterizing QPT,
 e.g., the fidelity as the overlap of ground states \cite{lee2006,zan2006,zan2007}
 and the LE \cite{quan2006,yuan2007,li2007,ros2007}.
 As shown in Ref.\cite{wgw2010}, in the neighborhood of the critical point of the Ising chain,
 an effective Planck constant can be introduced and the semiclassical theory is useful in
 predicting the decaying behavior of the survival probability, which is a special case of the LE.
 In this paper, we study the validity of the
 semiclassical prediction when the effective Planck constant is increased.

 The paper has the following structure.
 In Sec.\uppercase\expandafter{\romannumeral2}, we recall the semiclassical approach to the LE decay.
 In Sec.\uppercase\expandafter{\romannumeral3}, we study the LE decay in the deep-quantum region in
 the sawtooth model and in the kicked rotator model.
 Section \uppercase\expandafter{\romannumeral4} is devoted to a study of the
 LE decay in the vicinity of the QPT of the 1D Ising chain in a transverse field.
 Finally, conclusions and discussions are given in Sec.\ref{sect-con}.

 \section{Semiclassical approach to the LE}
 \label{ale}

 Before presenting our results, let us first recall semiclassical predictions for the decay of LE.
 As well known, the quantum transition amplitude from a point $\mathbf{r}_0$ to a point $\mathbf{r}$
 in a $d$-dimensional configuration space
 within a time period $t$ can be expressed in terms of Feynman's path integral \cite{fey1948,fey1965}.
 In the semiclassical limit, one may use the stationary phase approximation to approximately compute
 the transition amplitude.
 Contributions from paths close to classical trajectories
 give the following well-known semiclassical evolution, in terms of Van Vleck-Gutzwiller propagator
 $K_{sc}(\mathbf{r};\mathbf{r}_0;t)$:
 \be \label{Psi-sm}
 \Psi_{sc}(\mathbf{r};t)=\int d\mathbf{r}_0 K_{sc}(\mathbf{r};\mathbf{r}_0;t)\Psi_0(\mathbf{r}_0),
 \ee
 where $K_{sc}(\mathbf{r};\mathbf{r}_0;t) = \sum_sK_{s}(\mathbf{r};\mathbf{r}_0;t)$ and
 \be
 K_{s}(\mathbf{r};\mathbf{r}_0;t) = \frac{C_s^{1/2}}{(2\pi i\hbar)^{d/2}}
 \exp\left[\frac{i}{\hbar}S_s(\mathbf{r};\mathbf{r}_0;t)-\frac{i\pi}{2}\mu_s\right].
 \ee
 Here, the subscript $s$ indicates classical trajectories,
 $C_s^{1/2}=|\det(\partial^2S_s/\partial r_{i0}\partial r_j)|$, $\mu_s$ is the Maslov index
 counting conjugate points, and $S_s(\mathbf{r};\mathbf{r}_0;t)$ is the action, i.e., the time integral of the
 Lagrangian along the trajectory $s$, $S_s(\mathbf{r};\mathbf{r}_0;t) = \int_0^t dt' \mathcal{L}$.

 Let us consider an initial narrow Gaussian wave packet,
 \be
 \Psi_0(\mathbf{r}_0) = \Big( \frac{1}{\pi\xi^2} \Big) ^{d/4} \exp\Big[ \frac{i}{\hbar}\tilde{\mathbf{p}}_0
 \cdot \br_0 -\frac{(\mathbf{r}_0-\tilde{\mathbf{r}}_0)^2}{2\xi^2} \Big],
 \ee
 where $(\tilde{\mathbf{r}}_0,\tilde{\mathbf{p}}_0)$ indicates the packet center
 and $\xi$ is the dispersion.
 Semiclassically, the LE is written as $M_{sc}(t)=|m_{sc}(t)|^2$, where
 \be
 m_{sc}(t) = \int d \mathbf{r} \left [ \Psi^{H_1}_{sc}(\mathbf{r};t) \right ]^\ast
 \Psi^{H_0}_{sc}(\mathbf{r};t).
 \ee
 As shown in Refs.\cite{jal2001,van2003}, the amplitude $m_{sc}(t)$ has the following explicit expression,
 \be
 m_{sc}(t)\simeq \Big( \frac{\xi^2}{\pi\hbar^2}\Big)^{\frac{d}{2}}
 \int d\mathbf{p}_0 \exp\Big[ \frac{i}{\hbar} \Delta S(\mathbf{p}_0,\tilde{\mathbf{r}}_0;t)-
 \frac{(\mathbf{p}_0-\tilde{\mathbf{p}}_0)^2}{(\hbar/\xi)^2}\Big],
 \label{msc}
 \ee
 where $\Delta S(\mathbf{p}_0,\tilde{\mathbf{r}}_0;t)$
 is the action difference along two nearby trajectories in two systems $H_1$ and $H_0$.
 In the first-order classical perturbation theory,
 with the difference between the two trajectories neglected, one has
 \be
 \Delta S(\mathbf{p}_0,\tilde{\mathbf{r}}_0;t) \simeq \epsilon\int^t_0 dt'V[\mathbf{p}_0(t')].
 \label{ds}
 \ee

 The LE amplitude in Eq.(\ref{msc}) can be written as an integration over $\Delta S$. As a result,
 the LE is written as
 \be
 M(t) = \left| \int d\Delta S e^{i\Delta S/\hbar} P(\Delta S)\right|^2,
 \label{pdsf}
 \ee
 where $P(\Delta S)$ is the distribution of the action difference defined by
 \bey
 P(\Delta S) & \simeq & \left( \frac{\xi^2}{\pi\hbar^2}\right)^{\frac{d}{2}} \int d\mathbf{p}_0
 \delta[\Delta S - \Delta S(\mathbf{p}_0,\mathbf{\tilde{r}}_0;t)]\nonumber \\
 && \cdot \exp\left[-\frac{(\mathbf{p}_0-\tilde{\mathbf{p}}_0)^2}{(\hbar/\xi)^2}\right].
 \label{PDS}
 \eey

 Let us first discuss the LE decay in chaotic systems.
 In such a system, with an average performed over initial states,
 the distribution $P(\Delta S)$ is usually not far from a Gaussian distribution.
 When the perturbation is not strong, in the so-called FGR regime,
 deviation of  $P(\Delta S)$ from the Gaussian distribution can be neglected.
 In this case, the semiclassical theory predicts the following FGR decay for the LE \cite{cer2002},
 \be
 {M_{sc}}(t) \simeq e^{-2\sigma^2R(E)t},
 \label{fgrm}
 \ee
 where $\sigma=\epsilon / \hbar$ and $R(E)$ is the classical action diffusion constant,
 \be
 R(E) = \int^{\infty}_{0} dt \langle (V[r(t)]- \la V\ra)( V[r(0)]-\la V \ra) \rangle,
 \label{re}
 \ee
 with $\la  \cdot \ra$ indicating the average over the primitive periodic orbits of a very long period.
 Consistently, similar results for the FGR decay can also be obtained in the approach of random
 matrix theory\cite{jac2001,jac2002,cuc2002}.

 With increasing perturbation strength, deviation of the distribution $P(\Delta S)$ from the Gaussian
 form can not be neglected, and one enters into the so-called Lyapunov regime.
 In this regime, due to the quadratic dependence of the FGR rate on the perturbation strength,
 the part of the LE having the FGR decay decreases quite fast;
 beyond a time scale at which this part of the LE reduces to a negligible value \cite{foot1},
 the LE will be dominated by the contribution from the
 above-mentioned deviation of the distribution $P(\Delta S)$ from the Gaussian form \cite{wgw2004}.
 It has been found that the latter contribution is mainly given by
 $\Delta S$ close to its stationary points with respect to initial momentum, and
 the stationary phase approximation predicts
 the following perturbation-independent decay \cite{wgw2005,wgw2005t}:
 \be
 M_{sc}(t) \propto \exp\left[-\Lambda_1(t)t\right],
 \label{lym}
 \ee
 where
 \be
 \Lambda_1(t)=-\frac{1}{t} \lim_{\delta x(0)\rightarrow 0} \ln
 \overline{\Big|\frac{\delta x(t)}{\delta x(0)}\Big|^{-1}},
 \label{lbd}
 \ee
 with the average taken over initial states.
 One should note that $\Lambda_1(t)$ is usually not equal to the Lyapunov exponent $\lambda_L$,
 \be
 \lambda_L =\lim_{t\to\infty} \frac{1}{t} \lim_{\delta x(0)\rightarrow 0} \ln
 \Big|\frac{\delta x(t)}{\delta x(0)}\Big|,
 \label{lyapu}
 \ee
 due to local fluctuations.
 When the time $t$ is sufficiently long such that $\Lambda_1(t)$ becomes close to its long-time limit,
 the LE has a decay determined by the long-time limit
 of $\Lambda_1(t)$ discussed in Ref.\cite{sil2003}.
 In a system with a homogeneous phase space, i.e., with a constant local Lyapunov exponent,
 $\Lambda_1(t)$ is given by the Lyapunov exponent and the LE has the Lyapunov decay \cite{jal2001}.


 Next, we discuss the integrable case.
 In a 1D regular system with periodic classical motion,
 the LE has the following semiclassical expression up to a second-order perturbation
 contribution \cite{wan2007}:
 \be
 M_{sc}(t) \simeq {c_0}{(1+\xi^2 t^2)^{-1/2}}
 e^{ -\Gamma t^2 /(1+\xi^2 t^2)} ,
 \label{Mt-scr}
 \ee
 where $c_0\sim 1$ and
 \be \Gamma =  \frac{1}{2}\left( \frac{\varepsilon w_p}{\hbar} \frac{\partial U}{\partial p_0}  \right)^2, \ \
 \xi = \left| \frac{\epsilon w_p^2}{2\hbar} \frac{ \partial^2 U} { \partial p_0^2} \right|,
 \ee
 with the derivatives evaluated at the center of the initial Gaussian wave packet.
 Here, $w_p$ is the width of the initial Gaussian wave packet in the momentum space,
 $U= \frac 1{T_p} \int_0^{T_p} V\,dt$, and $T_p$ is the period of the classical motion.
 Equation (\ref{Mt-scr}) shows that the LE has an initial Gaussian decay predicted in
 Refs.\cite{pro2002t,pro2002} and a long-time power-law decay $1/t$.

 In the opposite case of a regular system with many degrees of freedom and many different frequencies,
 when the time is not long such that the classical motion does not show any sign of quasi-periodicity,
 the classical motion looks like a chaotic one.
 In this case, the LE also has an initial Gaussian decay \cite{pro2002t,pro2002},
 but it is followed by a FGR-type decay in Eq.(\ref{fgrm}) \cite{wgw2010}, with
 \be
 R(E) = \frac{1}{2t} \left( \left\langle \left[\int_0^{t} V(t') dt' \right]^2 \right\rangle
 -\left\langle\int_0^{t} V(t') dt'\right\rangle^2 \right).
 \label{re-2}
 \ee

 Finally, we note that Eq.(\ref{pdsf}) is a general expression,  not restricted to the
 case of the semiclassical limit.
 To show this point, one may use Feynman's path-integral formulation.
 For brevity, lets us write Feynman's propagator as
 \begin{equation}\label{FP}
 K_F(\br,\br_0;t) = {\cal N} \sum_{\alpha} \exp \left \{ i S_{\alpha}(\br,\br_0;t) / \hbar \right \},
 \end{equation}
 where $\alpha $ indicates possible paths going from $\br_0$ to $\br$ within a time interval
 $t$, and ${\cal N}$ is the normalization coefficient.
 Using this propagator, the exact time evolution of the wave function $\Psi(\br,t) $
 can be written in a form
 similar to that in Eq.(\ref{Psi-sm}), with $K_{sc}$ replaced by $K_F$.
 Then, substituting the expression obtained into the definition of $m(t)$ in Eq.(\ref{le}),
 one obtains
 \begin{equation}\label{}
 m(t) = \N \N' \int d \br_0 \br_0' \sum_{\alpha \alpha'} \exp (i \Delta S_e /\hbar) \Psi_0(\br_0)
 \Psi_0^*(\br_0'),
 \end{equation}
 where $ \Delta S_e = S_\alpha^{H_0} - S_{\alpha'}^{H_1}$.
 It is seen that the LE amplitude $m(t)$ can always be written as an integration over
 the exact action difference $\Delta S_e$, with the distribution $P(\Delta S)$ defined accordingly.
 As a result, the LE can always be written in the form of Eq.(\ref{pdsf}).

 \section{LE decay in the deep quantum region of two kicked systems}

 \begin{figure}
 \includegraphics[width=\columnwidth]{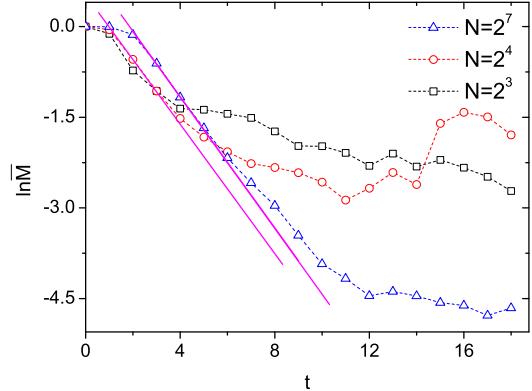}
 \caption{LE decay in the FGR regime in the sawtooth model,
 for different values $N$ of the dimension of the Hilbert space.
 Parameters: $K=2.0$ and $\sigma=0.5$.
 The LE (averaged over initial states) has two stages of decay.
 In the first stage, it follows the semiclassically-predicted FGR decay (solid straight lines), namely,
 $e^{-\sigma^2 \pi^4t /45}$ in Eq.(\ref{fgr-st}); in the second stage, it decays with slower rates.}
 \label{swFGR}
 \end{figure}

 In studying the validity of semiclassical predictions in the deep quantum region,
 it would be convenient to employ models in which effective Planck constants can be suitably introduced.
 In such a model, the value of the effective Planck constant gives a natural measure to the quantum ``deepness.''

 \subsection{Two kicked models}

 \begin{figure}
 \includegraphics[width=\columnwidth]{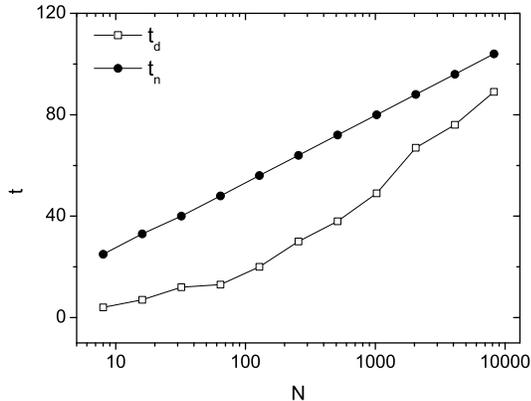}
 \caption{Variation of $t_d$ (empty squares)
 with $N$ in the sawtooth model, where $t_d$ is the time at which the second-stage
 decay of the LE in the FGR regime appears (see Fig.\ref{swFGR}).
 For comparison, we also plot $t_n$ (solid circles),
 the time at which the FGR decay is expected to reach the saturation value $1/N$,
 i.e.,  $t_n=(45\ln(N))/(\sigma^2\pi^4)$.
 Parameters: $K=2.0$, $\sigma=0.2$.
 }
 \label{fgrtc}
 \end{figure}

 We employ the sawtooth model and the kicked rotator model, whose Hamiltonians have the following form,
 \be
 H = \frac{1}{2}p^2 + V(r)\sum^{\infty}_{n=0}\delta(t-nT),
 \ee
 where $V(r) =- k(r-\pi)^2/2$ for the sawtooth model
 and $V(r) = k \cos r$ for the kicked rotator model.
 Here, for simplicity in the discussion, we consider their dimensionless form.
 Hereafter, we take the unit Planck constant, $\hbar =1$.

 The classical dynamics in the kicked rotator model generates the standard map,
 \bey
 \tilde{p}_{n+1}&=&\tilde{p}_n+K \sin(r_n)\ \ \ \ (\mathrm{mod} \ 2\pi),\nonumber\\
 r_{n+1}&=&r_n+\tilde{p}_{n+1}\ \ \ \ \ \ \ (\mathrm{mod} \ 2\pi),
 \eey
 where $\tilde{p}_n=T p_{n}$, $K=kT$.
 The classical motion is regular for sufficiently small $K$ and is almost chaotic for $K$
 larger than 6 or so.
 In the sawtooth model, one has the following classical mapping,
 \bey
 \tilde{p}_{n+1}&=&\tilde{p}_n+K (r_n-\pi)\ \ \ \ \ (\mathrm{mod} \ 2\pi),\nonumber\\
 r_{n+1}&=&r_n+p\tilde{}_{n+1}\ \ \ \ \ \ \ \ \ (\mathrm{mod} \ 2\pi).
 \label{stm}
 \eey
 Equation.(\ref{stm}) can be written in the matrix form
 \be
 \left(
   \begin{array}{c}
     \tilde{p}_{n+1} \\
     r_{n+1}-\pi \\
   \end{array}
 \right)
 =\left(
    \begin{array}{cc}
      1 & K \\
      1 & K+1 \\
    \end{array}
  \right)
  \left(
    \begin{array}{c}
      \tilde{p}_n \\
      r_n-\pi \\
    \end{array}
  \right).
 \ee
 The constant matrix in the above equation possesses two eigenvalues $1+(K\pm\sqrt{(K)^2+4K})/2$.
 The motion of the classical sawtooth model is chaotic for $K>0$,
 with the Lyapunov
 exponent \be\lambda_L = \ln(\{2+K+[(2+K)^2-4]^{1/2}\}/2), \label{ly}\ee
 given by the larger eigenvalue of the constant matrix.

 \begin{figure}
 \includegraphics[width=\columnwidth]{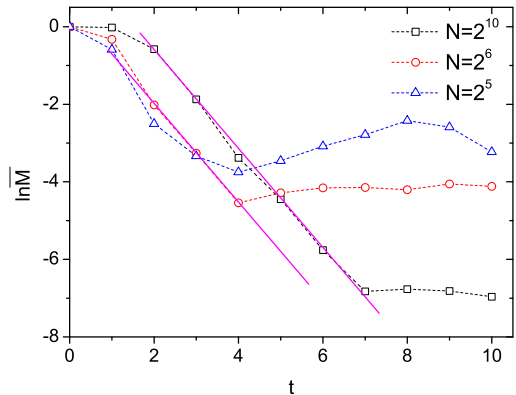}
 \caption{Similar to Fig. \ref{swFGR}, but for $\sigma=3.0$ in the Lyapunov regime
 of the sawtooth model.
 The LE has a decay close to the semiclassically-predicted Lyapunov decay, namely,
 $e^{-\lambda_L t}$ (solid lines), for $N \ge 2^6$.}
 \label{swLY}
 \end{figure}

 We utilize the method of quantization on a torus to get the quantum versions of the above two classical
 systems, with periodic boundary conditions for the coordinate and momentum variables,
 $0\leq r<r_m$, $0\leq p<p_m$
 \cite{han1980,man1991,wil1994,haa2001}.
 For a Hilbert space with dimension $N$, an effective Planck constant can be introduced,
 denoted by $h_{\rm eff}$, with $h_{\rm eff} = r_m p_m / N$.
 In the specific choice of $r_m=p_m=2\pi$, which will be taken in what follows,
 one has $h_{\rm eff} = (2\pi)^2/N$, hence, $\hbar_{\rm eff} = 2\pi/N$.
 The value of  $\hbar_{\rm eff}$ gives a measure to the ``deepness'' in the quantum region.
 The evolution operator for one period of time $T$, with $T=2\pi/N = \hbar_{\rm eff}$, is written as
 \be
 U=\exp\left[ -\frac{i}{2\hbar_{\mathrm{eff}}}\tilde{p}^2\right]\exp\left[ -\frac{i}{\hbar_{\mathrm{eff}}}
 \tilde{V}(r)\right],
 \ee
 where $\tilde{V}(r)=TV(r)$.

 In the two kicked models discussed above,
 the quantity $R(E)$ appearing in the FGR decay
 in Eq.(\ref{fgrm}) has the following expression \cite{lak1999,cer2000},
 \be \label{RE}
 R(E) = \frac{1}{2} C(0) + \sum^{\infty}_{l=1}C(l),
 \ee
 where $C(l) = \langle\{ V[r(l)] - \langle V\rangle \} \{ V[r(0)] - \langle V\rangle \} \rangle$.
 In the sawtooth model with an integer $K$, $C(0) = \pi^4/45$ and $C(l)=0$ for $l\neq0$,
 hence,
  \begin{equation}\label{fgr-st}
 {M}_{sc}(t) \simeq e^{-\pi^4\sigma^2t/45}.
 \end{equation}
 Meanwhile, in the kicked rotator model, $R(E)$ is a function of the parameter $K$ and does not have
 an explicit analytical expression.

 \begin{figure}
 \includegraphics[width=\columnwidth]{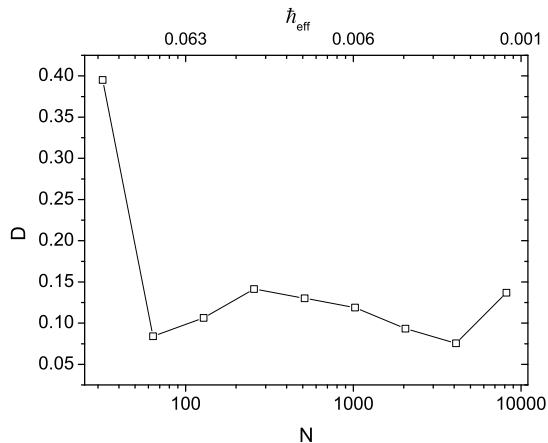}
 \caption{Variation of the deviation $D$ in Eq.(\ref{des}) with $N$ in the Lyapunov regime of the sawtooth model,
 with parameters $K=2.0$ and $\sigma=3.0$.
 The value of $D$ remains small for $N\geq N_c=64$ and becomes large when $N$ is smaller than $N_c$.}
 \label{lyva}
 \end{figure}

 \subsection{Numerical results in the sawtooth model}

 In this subsection, we discuss our numerical simulations obtained in the sawtooth model,
 with Gaussian wave packets as the initial states.
 In the FGR regime, when $N$ is not large,
 it was found that, beyond some initial times,  the LE has two stages of decay (see Fig.\ref{swFGR}):
 In the first stage, the LE follows the semiclassically predicted FGR decay,
 while, in the second stage, it is somewhat slower than the FGR decay.
 After these two stages of decay,
 the LE oscillates around its saturation value, which is on average approximately equal to $1/N$ \cite{pro2002}.
 Interestingly, the first-stage decay of the LE exists even for small dimension $N$ of the Hilbert space,
 in other words, for values of the effective Planck constant not much smaller
 than its upper border $\hbar_{\rm eff}^{\rm ub}=2\pi$.
 Hence, it exists in the deep quantum regime.

 Let us use $t_d$ to indicate the transition time of the above-discussed two stages of decay of the LE,
 i.e., the time at which an obvious deviation from the FGR decay appears.
 As seen in Fig.\ref{fgrtc}, $t_d$ increases with increasing $N$.
 In addition, we observe that, when $N$ is increased,
 the second-stage decay of the LE approaches the FGR decay,
 that is, the difference between the decay rates of the two stages decreases.

 To get a further understanding in the above-discussed first and second stages of decay of the LE,
 let us reconsider the expression of the LE in Eq.(\ref{pdsf}).
 As discussed previously, Eq.(\ref{pdsf}) is not just a semiclassical expression,
 but, is an exact expression, if the distribution
 $P(\Delta S)$ is appropriately defined in terms of contributions from Feynman paths.
 The distribution $P(\Delta S)$ always has some deviation from its Gaussian approximation,
 which we denote by $P_G$, with $G$ standing for Gaussian, that is,
  \begin{equation}\label{}
 P(\Delta S) = P_G + \Delta P.
 \end{equation}
 The above-discussed numerical results imply that the deviation $\Delta P$ is not sufficiently
 large for times shorter than $t_d$ (beyond some initial times).
 As a result, the LE still follows the FGR decay.

 However, for times beyond $t_d$, the deviation can not be neglected.
 In fact, the deviation $\Delta P$ has mainly two sources:
 One comes from contributions not included in the stationary phase
 approximation, which has been used when deriving the semiclassical propagator from Feynman's path
 integral formulation.
 The other is related to the fact that
 the right-hand side of Eq.(\ref{PDS}) for classical trajectories does not have an exact Gaussian form.
 The second-stage non-FGR decay of the LE appears for quite small values of the dimension $N$,
 which correspond to values of the effective Planck constant not much smaller than the upper border
 $\hbar_{\rm eff}^{\rm ub}$.
 This implies that its deviation from the FGR-decay might have a non-semiclassical origin, i.e.,
 the above-mentioned first factor might play the major role here.

 Next, we dicuss the Lyapunov regime in the sawtooth model.
 In this regime, we did not observe a two-stage decay similar to that discussed above in the FGR regime.
 For large $N$, as shown in previous work \cite{wgw2004},
 the LE has approximately the semiclassically predicted Lyapunov decay, with the decaying rate given by the Lyapunov
 exponent. (The sawtooth model has a homogeneous phase space.)
 With decreasing $N$, as shown in Fig.\ref{swLY}, the decay of the LE gradually deviates from the
 Lyapunov decay.

 \begin{figure}
 \includegraphics[width=\columnwidth]{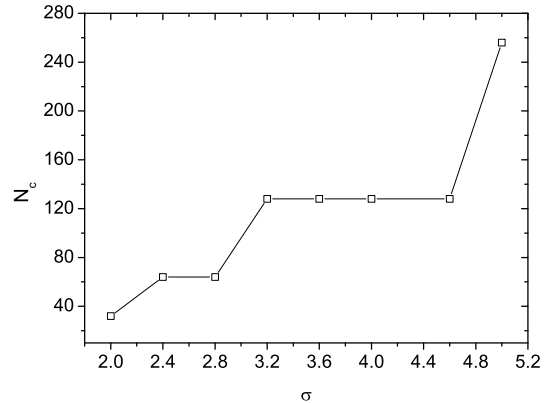}
 \caption{Variation of $N_c$ with $\sigma$ in the Lyapunov regime of the sawtooth model with
 parameter $K=1.0$.}
 \label{sigN}
 \end{figure}

 To quantitatively characterize the above-discussed deviation of the exact LE decay from
 the semiclassical prediction in the Lyapunov regime,
 we have studied the standard deviation of $x_n \equiv |\ln \overline{M}_e(t=n)-\ln {M}_{sc}(t=n)|$,
 where $ \overline{M}_e(t)$ denotes the exact numerical result.
 That is, we have studied the quantity $D$,
 \be
 D \equiv  \sqrt{\frac{1}{M} \sum^{M}_{n=1}(x_n-\overline{x})^2},
 \label{des}
 \ee
 where $\overline x =\frac{1}{M} \sum^{M}_{n=1}x_n$ is the average value of $x_n$.
 Our numerical simulations show
 that the value of $D$ remains small for large $N$ and becomes not small when $N$ is below
 some value, which we denote by $N_c$ (see Fig.\ref{lyva}).
 That is, the semiclassical prediction for the LE decay works well for $N$ above $N_c$, but,
 not well for $N$ below $N_c$.

 The value of $N_c$ was found to be dependent on the parameter $\sigma $, as shown in Fig.\ref{sigN}.
 On average, $N_c$ increases with increasing $\sigma$.
 This dependence may be related to a requirement used in the derivation of
 the above-mentioned semiclassical predictions for the LE decay, namely, $\epsilon$ being small.
 Indeed, due to the relation $\epsilon =\sigma \hbar_{\mathrm{eff}} = 2\pi \sigma /N $,
 to keep $\epsilon$ at a fixed small value, $N$ should be proportional to $\sigma$.

 \begin{figure}
 \includegraphics[width=\columnwidth]{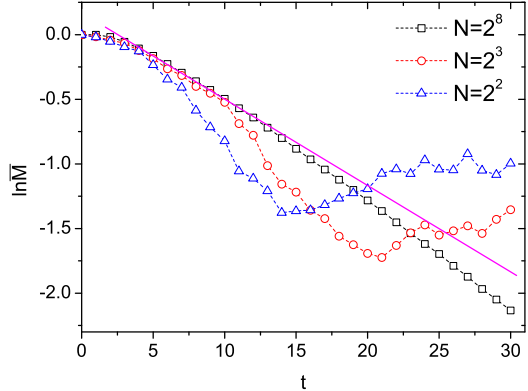}
 \caption{Same as in Fig. \ref{swFGR}, for the kicked rotator model
 with parameters $K=11.0$, $\sigma=0.3$ in the FGR regime.
 The semiclassical prediction of the FGR decay, $e^{-2\sigma^2R(E)t}$ with $R(E)=0.375$,
 is indicated by a solid line. }
 \label{krFGR}
 \end{figure}

 \subsection{Numerical results in the kicked rotator model}

 In the kicked rotator model, numerically we found the behaviors of the LE to be more or less similar to
 those in the sawtooth model discussed in the previous subsection, also with Gaussian wave packets
 as the initial states.
 Specifically, in the FGR regime, when $N$ is not large, we also observed a two-stage decay of the LE.
 But, in this model, the second-stage decay is faster than the first-stage FGR decay.
 See Fig.\ref{krFGR} for some examples,
 where the value of $R(E)$ in the FGR decay was computed numerically, making use of Eq.(\ref{RE}).

\begin{figure}
 \includegraphics[width=\columnwidth]{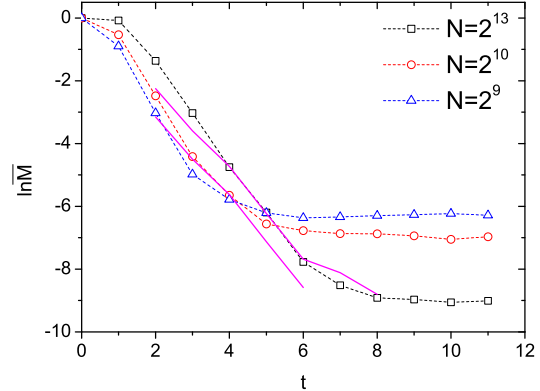}
 \caption{Similar to Fig.\ref{krFGR}, but for the Lyapunov regime
 with parameters $K=15.0$ and $\sigma=20.5$ in the kicked rotator model.
 The semiclassical prediction, given by Eq.(\ref{lym}), is indicated by solid lines.}
 \label{krLY}
 \end{figure}

 The kicked rotator does not have a homogeneous phase space. Hence, in the Lyapunov regime,
 the semiclassical prediction for the LE decay is not given by the Lyapunov exponent of the
 underlying classical dynamics, but, is given by Eq.(\ref{lym}).
 As expected, only for large $N$, did we numerically find agreement between the prediction of
 Eq.(\ref{lym}) and the exact LE decay beyond some initial time.
 Some examples are given in Fig.\ref{krLY}, where it is seen that
 the agreement is good for $N=2^{13}$ with  $t \in [4,8]$,
 while, the agreement is not good for $N\le 2^9$.
 In fact, for $N=2^9$, the LE approaches
 its saturation value before the semiclassically-predicted decay in Eq.(\ref{lym}) can be seen.

 \section{LE decay in the vicinity of a quantum phase transition}

 In this section, we discuss the LE decay in the vicinity of a critical point of
 a 1D Ising chain in a transverse field, with $N_p$ spins.
 As shown in Ref.\cite{wgw2010}, the semiclassical theory is useful in predicting the
 LE decay in the close neighborhood of
 those QPTs whose ground levels are infinitely degenerate at the critical points.
 The closer the controlling parameter $\lambda$ is to the critical point $\lambda_c$, the better
 the semiclassical theory might work.
 For this Ising chain, the semiclassical theory predicts an exponential decay for relatively long times.

 The Ising chain undergoes a QPT at the critical point in the thermodynamic limit $N_p\to \infty$.
 As will be discussed below, an effective Planck constant can be introduced in the low-energy region
 in this model, which is inversely proportional to $N_p$.
 We will be studying the extent to which the above-mentioned semiclassical prediction for an exponential
 decay of the LE may remain valid, when the value of $N_p$ is decreased.

 The dimensionless Hamiltonian of the 1D Ising chain is written as
 \be
 H(\lambda) = -\sum^{N_p}_{i=1}(\sigma_{i}^{z}\sigma_{i+1}^{z}+\lambda\sigma_{i}^{x}).
 \ee
 The spin-spin interaction intends to force the spins to polarize along the $z$ direction,
 while the transverse field intends to polarize them along the $x$ direction.
 Competition between the two interactions results in two critical points, $\lambda_c=\pm 1$,
 with the ferromagnetic phase for $-1<\lambda<1$ and the paramagnetic phase for $|\lambda|>1$.
 Without a loss of generality, we consider the critical point $\lambda_c=1$.

 \begin{figure}
 \includegraphics[width=\columnwidth]{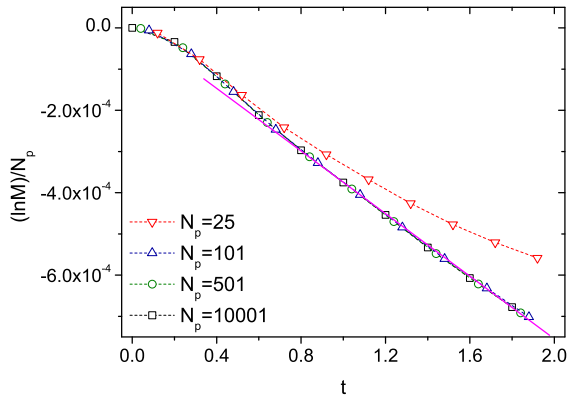}
 \caption{LE decay of the Ising chain for different values of $N_p$,
 with parameters $\lambda_0=\lambda_c-4\times10^{-2}$, $\lambda=\lambda_c-10^{-2}$ and $\lambda_c=1.0$.
 For large $N_p$ and relatively long times, the LE has an exponential decay as predicted by the semiclassical
 theory. (A solid straight line is drawn to guide the eyes.)
 Note also that the LE has a good scaling behavior of $\ln M \propto -N_p t$ for large $N_p$.
 For relatively small $N_p$ ($N_p=25$), the LE has neither the exponential decay nor the scaling behavior. }
 \label{istlnm2}
 \end{figure}

 The above Ising Hamiltonian can be diagonalized by utilizing the Jordan-Wigner and Bogoliubov
 transformations, giving \cite{lie1970,sac1999,pfe1970},
 \be
 H(\lambda) = \sum_{k} e_k(b_k^\dag b_k-1/2),
 \label{isdh}
 \ee
 where $b_k^\dag$ and $ b_k$ are fermionic creation and annihilation operators,
 $e_k$ is the  corresponding single quasi-particle energy,
 \be
 e_k = 2\sqrt{1+\lambda^2-2\lambda \cos(ka)},
 \label{isek}
 \ee
 and $k=2\pi m/aN_p$ with $m=-M,-M+1,\cdots, M$.
 Here, $a$ is the lattice spacing and $M=(N_p-1)/2$.

 \begin{figure}
 \includegraphics[width=\columnwidth]{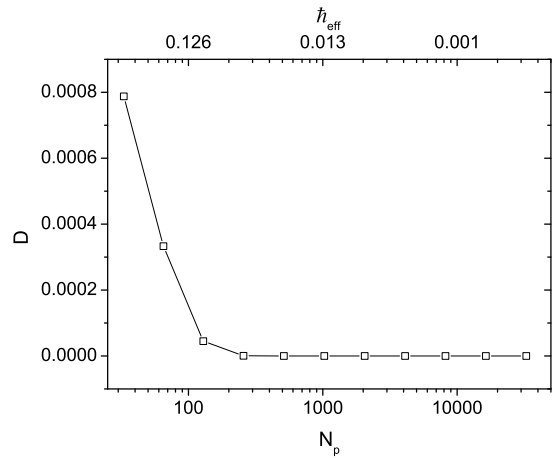}
 \caption{Variation of the deviation $D$ with the spin number $N_p$ in the Ising model,
 with parameters $\lambda_0=\lambda_c-4\times10^{-2}$ and $\lambda=\lambda_c-10^{-2}$.
 $D$ is large for $N_p< N_d \approx 100$. }
 \label{isva}
 \end{figure}

 As discussed in Ref.\cite{wgw2010}, in the very neighborhood of the critical point with $\lambda$
 sufficiently close to $\lambda_c$ and for sufficiently large $N_p$, the low-lying states have
 single-particle energies $e_k\approx (4\pi |m|)/N_p$ and
 can be mapped to bosonic modes by the method of bosonization \cite{sac1999}.
 A bosonic mode, labeled by $\alpha$, has a single-particle energy
 $e^b_\alpha \approx n_\alpha \delta E$, where $n_\alpha = 1,2,\ldots$ and $\delta E = 4\pi /N_p$.
 This expression of the single-particle energy $e^b_\alpha$ suggests that
 an effective Planck constant $\hbar_{\rm eff}$ may be introduced,
 \be
 \hbar_{\rm eff} = \delta E = 4\pi/N_p,
 \label{hefis}
 \ee
 which gives $e^b_\alpha = \hbar_{\rm eff}\omega_\alpha$, with $\omega_\alpha \approx n_\alpha $.
 In the case in which the frequencies $\omega_\alpha$ are sufficiently incommensurable, the classical
 counterpart has a motion like a chaotic one when the time is not long.
 Then, as discussed in Sec.II, the semiclassical theory predicts the exponential decay in Eq.(\ref{fgrm})
 with $R(E)$ given by Eq.(\ref{re-2}).

 \begin{figure}
 \includegraphics[width=\columnwidth]{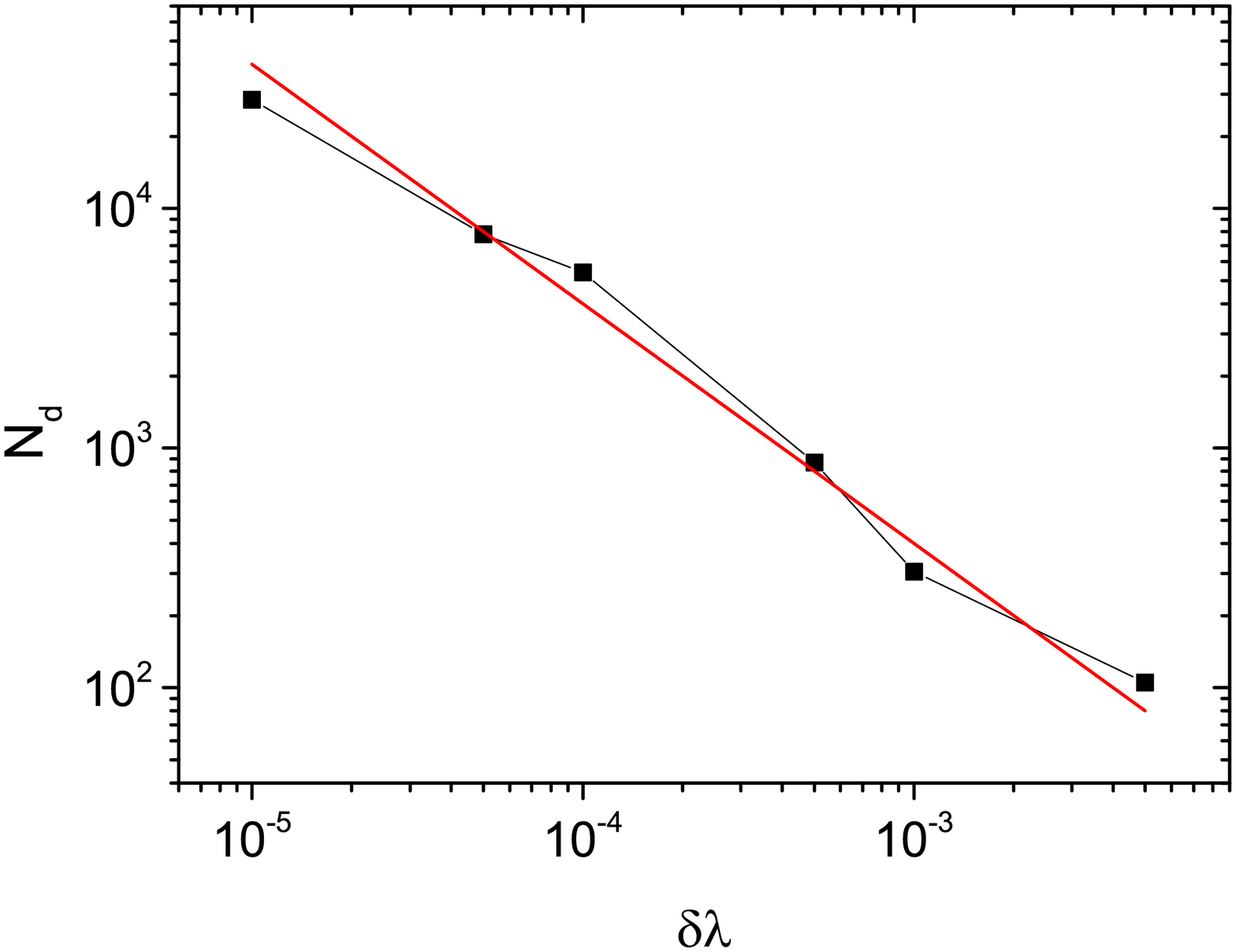}
 \caption{Variation of $N_d$ with the distance $\delta \lambda$ in the Ising model,
 with parameters $\lambda_0=\lambda_c-\delta \lambda$ and $\lambda=\lambda_0-\delta \lambda$.
 The solid line represents $N_d=2/(5\delta\lambda)$.}
 \label{npc}
 \end{figure}

 In computing the LE, $H_0$ in its definition in Eq.(\ref{le}) is taken as
 $H(\lambda_0)$ and $H_1$ as $H(\lambda)$.
 The initial state $|\Psi_0\rangle$ is chosen as the ground state of $H(\lambda_0)$;
 in this case, the LE is in fact a survival probability.
 Numerically, the LE was found to have an initial Gaussian decay, as predicted in Ref.\cite{quan2006}.
 For large values of $N_p$, the semiclassically-predicted exponential decay was also observed
 for relatively long times, i.e., beyond the initial Gaussian decay and before the revival time \cite{wgw2010}.
 However, when $N_p$ is decreased to some value, denoted by $N_d$,
 an obvious deviation from the exponential decay has been observed (see Fig.\ref{istlnm2}).

 Figure \ref{istlnm2} shows that for large $N_p$ the LE has a good scaling behavior,
 $\ln {M}(t) \sim N_p $.
 To understand this phenomenon, we note that here the perturbation $\epsilon V$ in the definition of the LE
 takes the form of $\epsilon = (\lambda_0 -\lambda)$ and
 \be
  V =  \sum^{N_p}_{i=1}  \sigma_{i}^{x}.
 \ee
 Then, according to Eq.(\ref{re-2}), the quantity $R(E)$ is given by the square of the summation of
 $N_p$ terms with mean zero, each of which is a time integration of
 $\epsilon (\sigma_{i}^{x} - \overline{\sigma_{i}^{x}})$.
 As discussed above, the classical counterpart has a motion like a chaotic one,
 hence, the time integrations mentioned above can usually be regarded as being uncorrelated.
 As a result, for large $N_p$, the quantity $R(E)$ is approximately proportional to $N_p$, hence
 $\ln {M}(t) \sim N_p $.

 To see more clearly the process of the above-discussed deviation of the LE
 from the semiclassically predicted exponential decay,
 we have calculated the deviation $D$ in Eq.(\ref{des}) for ${x_t=|\ln \overline{M}_e(t)
 -\ln {M}_{sc}(t)|/N_p}$ (see Fig.\ref{isva}).
 In our computation, $(\ln M_{sc})/N_p$ was computed in the large-$N_p$ limit.
 It is seen in  Fig.\ref{isva} that an obvious deviation from the semiclassically-predicted exponential
 decay appears at  $N_d \approx 100$.

 Furthermore, we found that the value of $N_d$ has a strong dependence on
 $\delta \lambda =\lambda -\lambda_c $,  as shown in Fig.\ref{npc}.
 Specifically, $N_d$ is almost inversely proportional to $\delta \lambda$.
 Therefore, the value of $N_d$ can be not large for $\delta \lambda$ not very small.
 However, for quite small $\delta \lambda$, $N_d$ can be very large.
 Since a large value of $N_d$ implies `deep' in the semiclassical regime,
 it is reasonable to expect that this deviation from the semiclassically predicted
 exponential decay may be due to the invalidity of some approximation used in the semiclassical derivation.

 Indeed, as shown below, the above-mentioned deviation can be explained by approximate commensurability
 of the frequencies $\omega_\alpha$, which may invalidate the derivation for the exponential decay.
 Let us go back to the single-particle energy $e_k$ in Eq.(\ref{isek}) and get its approximate expression
 for large $N_p$ and small $|m|$, with the $\lambda$-dependence written explicitly,
 \be
 e_k \simeq \frac{4\pi}{N_p} |m| \sqrt{\lambda +G^2_\lambda},
 \label{ek}
 \ee
 where
\begin{equation}\label{}
   G_\lambda = \frac{N_p \delta \lambda}{2\pi m }.
\end{equation}
 Note that for $\lambda=\lambda_c=1$, this expression gives the approximation
 used previously, namely, $e_k\approx (4\pi |m|)/N_p$.
 When the term $G_\lambda^2$ is small compared with 1, one can argue that the low-lying states of the model
 can still be mapped to bosonic modes.
 For $\lambda$ close to $\lambda_c$, the frequencies of the bosonic modes are written as
\begin{equation}\label{}
 \omega_\alpha \simeq n_\alpha\sqrt{1+G^2_\lambda}.
\end{equation}
 For the LE to have FGR-type exponential decay, $\omega_\alpha$ should be sufficiently incommensurable.
 Hence, the term $G_\lambda^2$ can not be very small, i.e., $G_\lambda$ should be larger than
 some small quantity.
 Obviously, the breakdown dimension $N_d$ estimated in this way is inversely proportional to
 $\delta \lambda $, in agreement with numerical results given in Fig. \ref{npc}. \\

 \section{Conclusions and discussions}
 \label{sect-con}

 In this paper, we have studied the change from validity to breakdown of some semiclassical predictions for the
 LE decay in several models, when the effective Planck constants are increased and the systems move from the
 semiclassical region to the deep quantum region.
 Our numerical results show that some semiclassical predictions for the LE decay work well even in the deep
 quantum region.

 In particular, in the FGR regime with intermediate perturbation strength
 in the two quantum chaotic systems studied,
 there is always some time interval within which the LE follows the FGR decay;
 the length of this time interval decreases when the effective Planck constant is increased.
 Making use of an exact expression of the LE, which is obtained resorting to Feynman's path integral
 formulation of quantum mechanics, it is argued that
 this phenomenon should be universal for quantum chaotic systems.
 This is in agreement with the fact that the same FGR decay can also be derived
 by other methods, namely, by the random matrix theory \cite{jac2001}
 and by a linear response theory \cite{pro2002t,pro2002}.
 Still in the FGR regime, beyond the time interval discussed above,
 deviation of the LE from the FGR decay has been observed in the two chaotic systems in the deep quantum region.
 This deviation is expected to be induced by non-semiclassical contributions
 and may also appear in other chaotic models.

 In the Lyapunov regime with relatively stronge perturbation, a different situation has been found.
 In particular, the semiclassical prediction has been found to be invalid in a sufficiently-deep quantum region.
 This difference from the FGR regime is understandable, since the mechanism for the LE decay is different
 in the two regimes.

 \acknowledgements

 This work was partially supported by the Natural Science Foundation of China
 under Grant Nos.~11275179 and 10975123 and the National Key Basic Research Program of China under Grant
 No.2013CB921800.

  \end{document}